# Pigment Network Detection and Classification in Dermoscopic Images Using Directional Imaging Algorithms and Convolutional Neural Networks

M. A. Rasel[1], Sameem Abdul Kareem[1], Unaizah Obaidellah[1, *]

[1]Department of Artificial Intelligence, Faculty of Computer Science and Information Technology, Universiti Malaya, Kuala Lumpur, 50603, Malaysia
[*]Corresponding Author: Unaizah Obaidellah. Email: unaizah@um.edu.my

**ABSTRACT**
Early diagnosis of melanoma, which can save thousands of lives, relies heavily on the analysis of dermoscopic images. One crucial diagnostic criterion is the identification of unusual pigment network (PN). However, distinguishing between regular (typical) and irregular (atypical) PN is challenging. This study aims to automate the PN detection process using a directional imaging algorithm and classify PN types using machine learning classifiers. The directional imaging algorithm incorporates Principal Component Analysis (PCA), contrast enhancement, filtering, and noise reduction. Applied to the PH2 dataset, this algorithm achieved a 96% success rate, which increased to 100% after pixel intensity adjustments. We created a new dataset containing only PN images from these results. We then employed two classifiers, Convolutional Neural Network (CNN) and Bag of Features (BoF), to categorize PN into atypical and typical classes. Given the limited dataset of 200 images, a simple and effective CNN was designed, featuring two convolutional layers and two batch normalization layers. The proposed CNN achieved 90% accuracy, 90% sensitivity, and 89% specificity. When compared to state-of-the-art methods, our CNN demonstrated superior performance. Our study highlights the potential of the proposed CNN model for effective PN classification, suggesting future research should focus on expanding datasets and incorporating additional dermatological features to further enhance melanoma diagnosis.

## 1. INTRODUCTION

Melanoma is a form of skin cancer that begins in the cells (melanocytes) that control the pigment in our skin [1]. This can be "metastasized" (the spread of cancer cells to other body parts) [2]. Early diagnosis is the only way to prevent this fatal disease. Observation of the lesion's characteristics such as color, shape, symmetry, and pigment network are key steps for early melanoma detection. Pigment network (PN) is a skin condition identified as an irregular fine network or netlike structure made of intersecting brown lines [3]. The PN of a lesion helps to spot abnormalities in a dermoscopic structure, providing information necessary to diagnose melanoma. Several approaches for studying dermoscopic images and detecting PN have been developed; however, only a few have addressed the classification of PN, which determines the abnormality status of the skin. The color, directionality, and topological features of the PNs are used clinically to classify them as an atypical pigment network (APN) and typical pigment network (TPN). The nature of grid distribution and color are the primary differences needed to distinguish between APN and TPN. The irregular distribution of an APN is usually characterized by dark-brown, gray, or black colors [4]. In contrast, a TPN has a light-brown structure that is evenly spread [5]. Atypical or dysplastic (abnormal or atypical appearance of a cell or tissue) naevi and certain melanomas exhibit an APN on the lesion [3]. These properties of APN and TPN are key for distinguishing melanocytic lesions from non-melanocytic lesions [6]. Fig. 1 shows a lesion with TPN and a lesion with APN taken from a publicly available dataset [7].

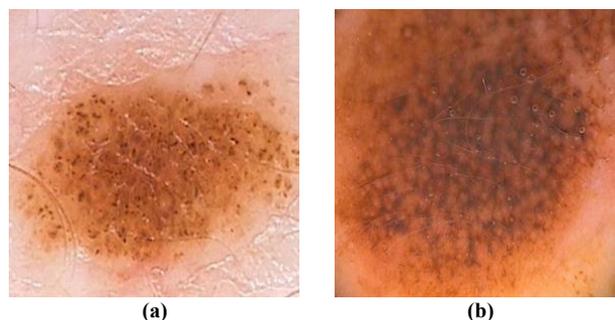

(a) (b)
Fig. 1: (a) is TPN with evenly spread light-brown color, and (b) is APN with dark-brown color.

The APN is one of the key factors identified in the 3-point checklist (asymmetry, APN, and blue-white structures) [8] and 7-point checklist (APN, blue-whitish veil, atypical vascular pattern, irregular streaks, irregular dots/globules, irregular blotches, and regression structures) [9] melanoma skin cancer diagnosis methods. The definition of an APN in the 3-point checklist method is a PN with irregular holes and thick lines. In the 7-point checklist method, APN means reticular lines, heterogeneous color-thickness, and asymmetric distribution within the lesion [10]. That is one of seven features with 2 points (melanoma ≥ 3 points) [10]. Another method to diagnose melanoma is called CASH (color, architecture, symmetry, and homogeneity) [11], where the PN is included in the homogeneity criteria. In dermoscopic images, APN is difficult to spot, typically requiring the expertise of a dermatologist to notice the anomaly during diagnosis. The use of computer analysis and image processing to support quantitative medical diagnosis is a powerful tool. As a result, computer-based approaches for analyzing dermatological images are sought after. An automatic system can objectify, speed up, and increase the efficiency of PN detection and classification, leading to more precise melanoma diagnosis.

Medical image processing involves the use of computational techniques to analyze and interpret medical images and plays a critical role in enhancing image quality, segmenting regions of interest (ROI), and extracting features necessary for diagnosis. Techniques such as Principal Component Analysis (PCA) [12], contrast enhancement [13], and noise reduction [14] are commonly employed to improve the accuracy of image interpretation. Convolutional Neural Networks (CNNs) have emerged as an essential tool for deep learning, particularly suited for medical image recognition [15, 16]. CNNs are capable of automatically learning features from raw image data, making them highly effective for image classification tasks. Their architecture, which includes convolutional layers, pooling layers, and fully connected layers, allows them to detect complex patterns and structures within images. Other machine learning classifiers, such as the Bag of Features (BoF) [17], are also used to classify images based on extracted features. These classifiers provide a robust framework for differentiating between typical and atypical PNs, aiding in the early detection of melanoma. By leveraging these advanced image processing techniques and machine learning classifiers, we aim to develop an automated approach for detecting and classifying PNs in dermoscopic images, thereby improving the accuracy and efficiency of melanoma diagnosis.

## 2. RELATED WORK

In this section, a review of similar studies with focus on different approaches to analyze PN on skin lesion images. A. Naser et al [18] introduced a technique to extract PN information for diagnosing skin cancer by investigating whether a lesion is normal or abnormal. This work was divided into four steps: artifacts elimination, PN extraction, feature extraction and lesion classification. The directional Gabor filters for detecting PN and artificial neural networks (ANN) for classification purposes were used in their methodology. S. Pathan et al. [19] proposed a methodology that includes artifacts detection, PN enhancement, PN classification, and lesion classification steps in detecting Melanoma. The blue channels of input RGB images were applied to detect hair and the 2D Gabor filters were adopted to highlight the PN. The extracted information of PN such as color, geometry and texture were analyzed for lesion classification using Support Vector Machine (SVM) and ANN. S. Maryam et al. [20] presented a PN detection idea in dermoscopic images using graphs technique for classifying moles and detecting Melanoma. The Laplacian of Gaussian filter was used after pre-processing to detect the sharp changes of intensity. The edge detection process was binary, then converted into a graph to find meshes or cyclic structures of the region of interest (ROI). Finally, noise or undesired cycles were eliminated from the graph of PN. Then the created images from the graph were applied to classify lesions. C. Barata et al. [21] proposed a direction approach for detecting PN in dermoscopic images, which comprises hair detection, reflection detection, image impainting and network detection process. A bank of directional filters was used to detect the line color, geometrical shape and topology of the PN for the dermoscopic image classification into two classes: with and without PN. G. Arroyo et al. [22] invented an algorithm to detect the reticular pattern of PN in lesion without any preprocessing and segmentation techniques, containing two blocks. Firstly, a supervised machine learning process was carried out to apply over an image to generate a mask with pixels candidates for being the part of PN. Secondly, the mask was applied to conduct a structural analysis process to find out whether ROI contained the PN or not. K. Kropidlowski et al. [23] developed a histogram-based method to detect APN and irregular streaks on the lesion combined with correction of image illumination, segmentation, feature extraction and classification. The classification process was done by a two-layer ANN to process four features such as: solidity, convexity, entropy, and contours distance-based skewness. K. Eltayef et al. [24] implemented a PN detection method by following four steps which were artifacts detection (reflection and hair), PN detection, feature extraction and lesion classification (normal or abnormal skin). The Gabor filter was united with the Sobel filter for artifacts detection, while Barata's method was employed to isolate the PN before, a two-layer feed-forward network was applied for a classification task.

The discussion of these methods serves as a guide for evaluating dermoscopic images. One of the dominant criteria for the 3-point checklist, 7-point checklist, and CASH methods is focused on both detecting the PN and classifying them into typical and atypical patterns. Tiny dense pigment rings (the grid) make the PN classification process a challenging task. This skin-gird line is difficult to be discerned with naked human eyes. Consequently, a less significant number of algorithms are developed to identify PN patterns based on colors, structures, and topological properties. Besides, to our knowledge, none of the algorithms isolated the PN from the lesion or ROI. Additionally, none of the above-mentioned works included a deep learning-based PN classification. Using less variation (only one source/dataset) of dermoscopic images was observed in those strategies, thereby the classification process of the PN is not feasible. APNs are a symptom of abnormality in a skin lesion that cannot be used alone to diagnose a disease like melanoma. Other factors, such as asymmetrical lesions and blue-white structures, play a role in melanoma diagnosis. Hence, isolating, feature extracting, and classifying PN into APN and TPN will serve as a starting point to facilitate melanoma detection.

Considering these gaps, the goals of the proposed work are- to detect and isolate PN on the dermoscopic images by implementing a directed imaging algorithm (which should be flexible to use on datasets of different origins), and classification of these isolated PN into the category of APN and TPN using classifiers. After that, the performance of the proposed approach over the other state-of-the-art approaches will be demonstrated.

## 3. METHODS

### 3.1 The proposed approach

This PN detection process was constructed by following several steps as illustrated in Fig. 2. The dermoscopic images were taken as input for this process. The source of dermoscopic images was the publicly available annotated dataset PH2 [7]. In addition, imaging directional filters (a combination of several directional imaging functions) were applied on the input image to isolate PN from the skin lesion to determine the condition (healthy or unhealthy) of ROI. A new dataset (output of imaging filters) was created by storing these isolated PN. Finally, new dataset and old annotated dataset were both trained separately to see the difference in classification results between images that have been used with directional filters and those without directional filters. Different classifiers such as SVM, ANN, CNN, and Bag of Features (BoF) were engaged to gauge the best classifiers for the PN classification. Next, to demonstrate the robustness of proposed approach, a comparative study was done with the result of existing works from the literature.

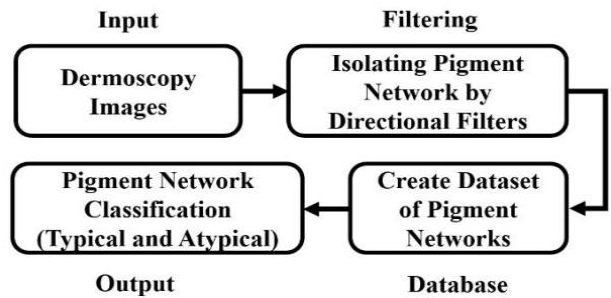

**Fig. 2: Overview of the proposed PN detection and classification system.**

### 3.2 Image acquisition

**With ground truth:** A publicly available online dataset (known as PH2) of 200 dermoscopic images (83 with TPN and 117 with APN) from the Hospital Pedro Hispano, Matosinhos database [7] was employed for the proposed directional algorithm. These RGB skin lesion images were captured during clinical exams using a dermoscopic device at a magnification of 20 times and saved in BMP and JPEG formats. They have an average resolution of 765x573 pixels.

**Without ground truth:** From the International Skin Imaging Collaboration (ISIC) archive another two datasets such as the ISIC 2019 [25, 26, 27], and the ISIC 2020 [28] were also included. The initial focus of these datasets was on dermoscopic images of individual skin lesions. The high-quality images of these two datasets have no specific pixel's resolution. To show the flexibility of the proposed approach on different origin datasets, a non-dermoscopic (captured images without a dermatoscope device) dataset called Med-Node [29] had been employed in the PN detection process. A total of 100 naevus and 70 melanoma images from the digital image archive of the Department of Dermatology of the University Medical Center Groningen was used for the development and testing of the Med-Node system for skin cancer detection from macroscopic images.

### 3.3 Isolating pigment network by directional filters

To detect the PN in lesion images, a comprehensive sequence of image processing algorithms was employed. The process begins with acquiring the input image in RGB format, followed by resizing to standardize dimensions. The image is then converted from RGB to the L*A*B color space to enhance color differentiation. The L*A*B channels are concatenated and reshaped for further analysis. Principal Component Analysis (PCA) is applied to reduce dimensionality and highlight key features. The transformed data is converted to grayscale to simplify processing. An enhanced image is produced, and a 10x10 filter is applied to smooth the image and reduce noise. Thresholding is performed to segment the image, followed by binary conversion to distinguish the pigment network from the background. Noise reduction techniques remove small pixel clusters to improve clarity. When the image is enhanced and filtered, noise elements and the significant features are more distinguishable. If noise reduction is applied too early, essential details might be lost because the algorithm might mistake small and important features for noise. The resulting image is complemented to invert pixel values, making the network more prominent. Finally, a color function is applied to enhance visualization, allowing for the effective separation of the PN from lesion images. Fig. 3 shows those consecutive steps of the proposed work which are available in the following sections.

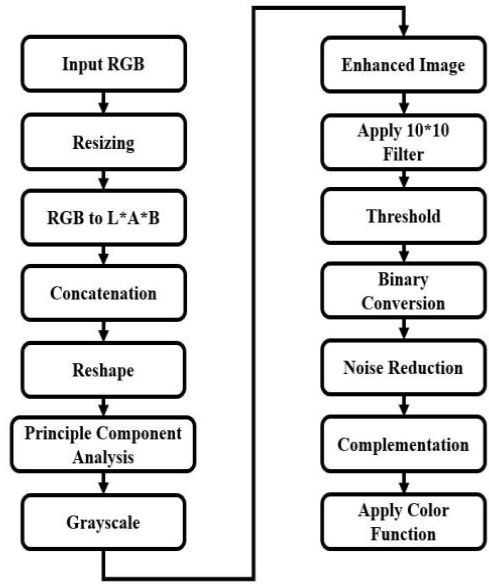

Fig. 3: The directional imaging approaches for detecting PN.

### 3.3.1 Initial Conversion Approaches
At first, the input images were resized into 512x512. The dimensions are in pixels as decimal dimensions are not accepted. Resizing was performed due to different image sizes of the datasets. Fig. 4 shows the resized input images.

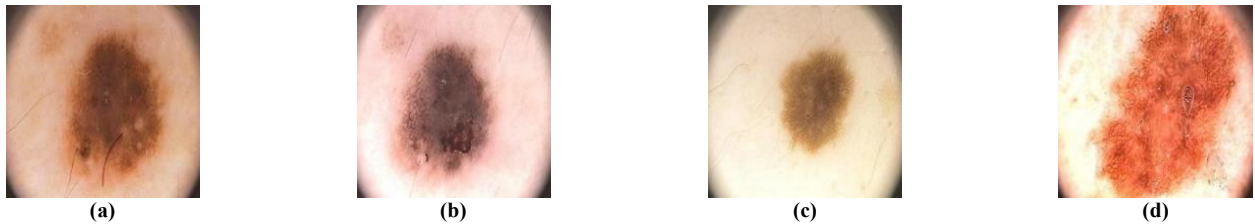

(a)            (b)            (c)            (d)

**Fig. 4: (a), (b), (c) and (d) are the outputs upon resizing input images.**

Subsequently, the images were converted into double precision. Image can be a grayscale intensity image, a true color image, or a binary image. Here, the image was an RGB image requiring further conversion of RGB values to the CIE (International Commission on Illumination) 1976 [30] L*a*b* values. According to this color space, the "L" represents the likeness of the color. The likeliness ranges from 0 to 100, where 0 is black and 100 is white. The "a" represents green to red, ranging from negative to positive. The higher the negative value, the brighter the green color, while the higher the positive value, the brighter the red color. The "b" represents blue to yellow, also ranges from negative to positive. Also, HSV [31] was used to compare with the L*a*b. To concatenate the (1, 0, 0) vector with the 3-dimensional arrays (since L*a*b and HSV have 3 channels), a function was created. Then element-wise binary operation was performed between that concatenated function and the last converted L*a*b and HSV images. Converting images from RGB to other color spaces like L*a*b or HSV is done to take advantage of the properties of these color spaces that can be more suitable for specific image processing tasks. Since perceptual uniformity and better color differentiation were required to detect PN, HSV color space was not suitable for this detection process. Next, the element-wise binary operation's output image was reshaped. For this process, the new dimension number should be null since it was not required here. Instead, 3 should be used since it was the existing dimension number of the last function's output.

### 3.3.2 PCA
The pre-processed images were next subjected to Principal Component Analysis (PCA). PCA is a mathematical procedure for reducing the number of dimensions in data [12], which involves- data standardization (a single image); calculation of covariance matrix for the features in image; calculation of eigenvalues and eigenvectors for covariance matrix; sorting eigenvalues and their corresponding eigenvectors; picking *k* eigenvalues and form a matrix of eigenvectors; and transforming the original matrix. This process returned the coefficients and the score of PCA. Then the scores were resized based on the size of the L*a*b image. Then, all the rows and the columns of the first channel were counted and stored. Next, the scores of PCA were subtracted by the minimum value of scores, and then the subtracted result was divided by the difference between the maximum and minimum value of scores. This division process was element-wise division. The final output of this analysis process was a grayscale image which was shown in Fig. 5 for the multiple input images.

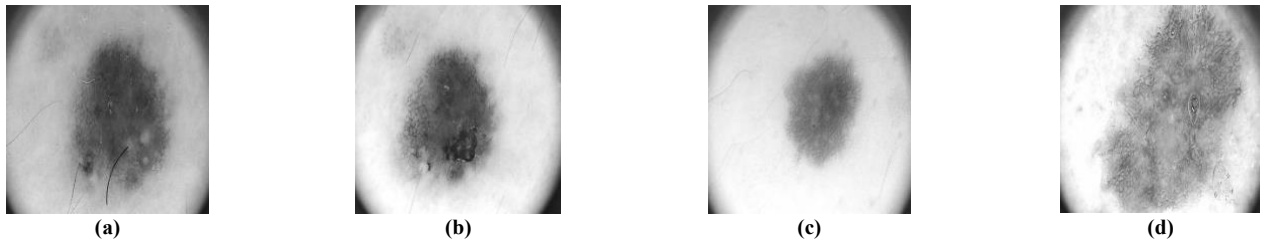

Fig. 5: (a), (b), (c) and (d) are the outputs after PCA for different input images.

### 3.3.3 Contrast Enhancement

Contrast enhancement was requisite for the last output grayscale image. This was done by employing the adaptive histogram equalization (AHE) [32] function. It improved the contrast of a grayscale image by applying contrast-limited AHE to alter the values [32]. The adaptive approach differed from the traditional histogram equalization in that it computes numerous histograms, each corresponding to a different portion of the image, and used them to disperse the image's brightness values. Therefore, it was ideal for enhancing local contrast and sharpening edge definitions in different parts of an image. This enhancement process was done block-by-block where the block size was 8x8 and the number of beams was 128. Fig. 6 presents the enhancing process outputs for AHE. Fig. 7 presents the enhancing process outputs for the traditional histogram equalization.

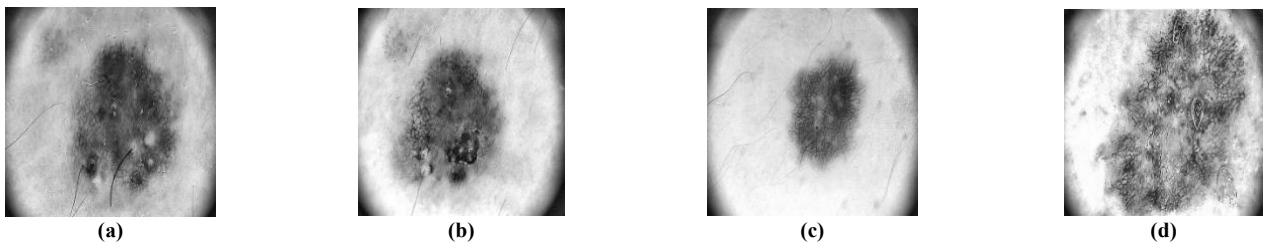

Fig. 6: (a), (b), (c) and (d) are the outputs after applying the AHE.

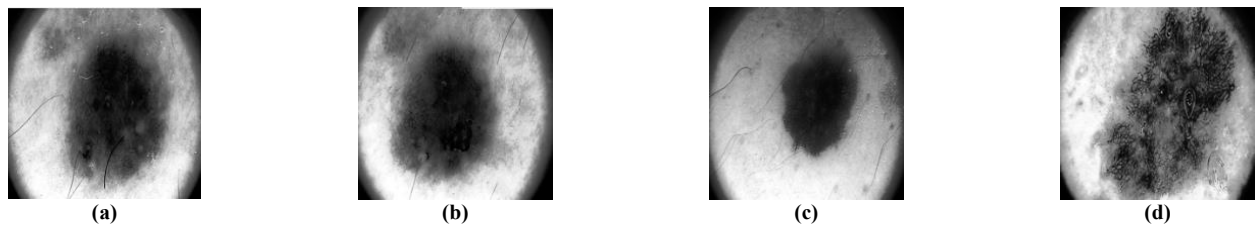

Fig. 7: (a), (b), (c) and (d) are the outputs after applying the traditional histogram equalization.

Then a 10x10 filter and Gaussian filter [33] were applied to the AHE-enhanced image. Fig. 8 and Fig. 9 show the output after applying the 10x10 and Gaussian filters on the AHE-enhanced images. Later, the enhanced images were subtracted from the 10x10 and Gaussian filtered images.

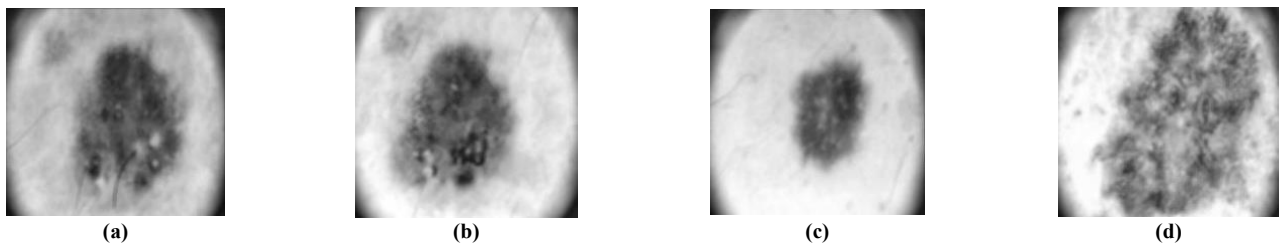

Fig. 8: (a), (b), (c) and (d) are the outputs after applying the 10*10 filter on the enhanced images.

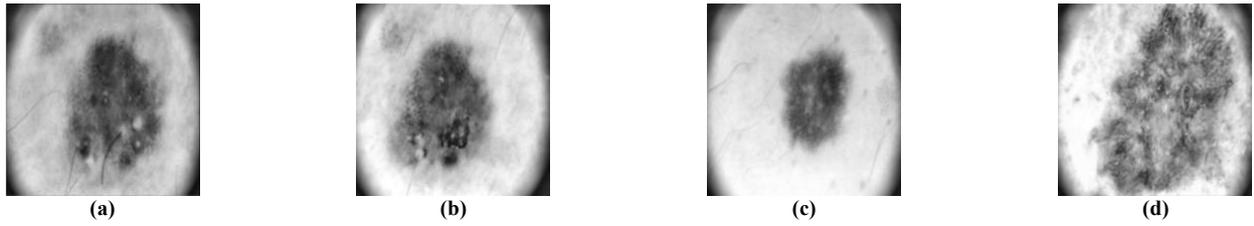

(a)  (b)  (c)  (d)

**Fig. 9: (a), (b), (c) and (d) are the outputs after applying the Gaussian filter on the enhanced images.**

### *3.3.4 Threshold Level Calculation*
The threshold level calculation was done on the subtracted images (from previous section 3.3.3: a 10*10 and Gaussian filtered images- Fig. 8 and Fig. 9) as a prerequisite to segment PN. In this calculation process, a function was created where an image was converted into an 8-bits data type; followed by counting the histogram and the number of beams. Also, the cumulative sum of the histogram count was calculated. The mean below and above 'T' was needed to find. Mathematically can be presented as-

$$T(i) = \frac{sum\ of\ (number\ of\ Bin * Histogram\ count)}{cumulative\ sum\ at\ the\ end} \quad (1)$$

Here, 'T' was the ratio of the sum of multiplication of bin number and histogram count to the cumulative sum indexed endmost. Then, to find the mean above 'T' and the mean below 'T', the cumulative sum of the histogram counts from 1 to $T(i)$ ('$i$' was an initial variable) was needed. Finally, the threshold was identified by calculating the average mean below 'T' and the mean above 'T'. Then a condition was given to make the absolute value of 'T' greater than 1. After that the threshold was normalized. Then, the created function was applied, and used the subtracted images (from previous section: 3.3.3) to return threshold level. Next, the subtracted images were converted to a binary image using threshold level. Instead of using threshold level directly, the value was reduced a little (i.e., 0.008) to make the PN more distinct and precise.

### *3.3.5 Reducing Noise and Colorization*
In this section, the PN has been converted to a binary format from a lesion image. In that binary image, pixels are typically either black (0) or white (1), representing the presence or absence of the PN, respectively. Here, the random variations in pixel intensity that do not represent the actual structures or patterns of PN. This type of noises can obscure or distort the true features in the image, making it harder to analyze. The connected components (clusters of contiguous pixels in the binary image) that consist of fewer than 100 pixels were considered noise. Each connected component represents a group of pixels that are adjacent and share the same value. This section involves eliminating those small connected components (i.e., noise) from the image. By removing these smaller clusters, the image becomes cleaner and more focused on the significant structures, which are the larger, more relevant parts of the PN. After subtracting, calculating threshold level, and reducing noise, Fig. 10 shows the output which can be considered as a clean image of PN from Fig. 8, and Fig. 11 shows the output which can be considered as a clean image of PN from Fig. 9. Among the Fig. 10 and Fig. 11, visually we understood that Fig. 10 has much more clean images than Fig. 11. For this reason, our next imaging technique was applied on the images of Fig. 10 which is coming from the 10x10 filtered-images.

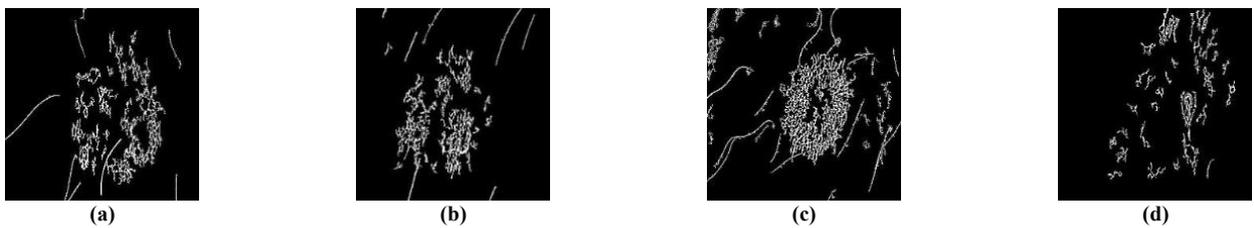

(a)  (b)  (c)  (d)

**Fig. 10: (a), (b), (c) and (d) are the noise reduction process outputs of the 10x10 filtered-images.**

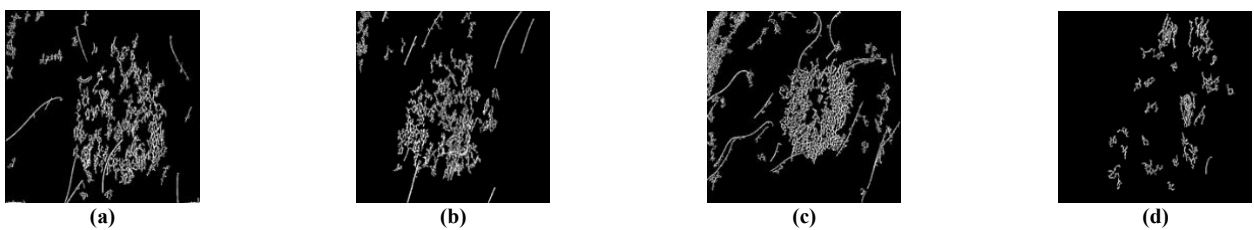

(a)  (b)  (c)  (d)

**Fig. 11: (a), (b), (c) and (d) are the noise reduction process outputs of the Gaussian filtered-images.**

Since these were binary images, they were converted into color images. The complement of the binary image was taken using the complementation process which returned another complemented image. This process mainly helped to improve the contrast. The image complement process computes the complement of a binary or intensity image. For binary images, the process replaced pixel values equal to 0 with 1 and pixel values equal to 1 with 0. For an intensity image, the process subtracts each pixel value from the maximum value that could be represented by the input data type and outputs the difference. As an illustration, consider the input pixel values given by m($i$) and the output pixel values given by n($i$). If the data type of the input was a double or single precision floating-point, the block outputs n($i$) = 1.0-m($i$). If the input was an 8-bit unsigned integer, the block outputs n($i$) = 255-m($i$). Fig. 12 shows this process outputs.

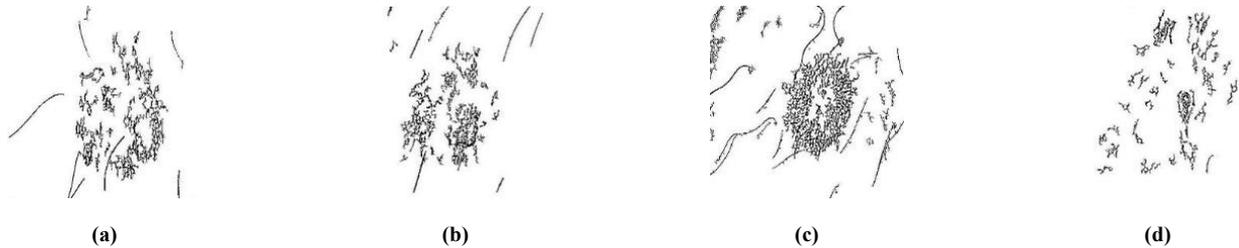

(a) (b) (c) (d)

**Fig. 12: (a), (b), (c) and (d) are the complementation process outputs.**

A function called colorized was created to color the last output image of PN without any difficulty. A new function was used to return the number of the input argument of the function. If this new function value was less than 3, the system would take the default color. After that, the complemented image was formed in a logical image. Then the earlier resized image and the color space were converted into 8-bits data type. If the dimension of the converted resized image was 2, then it should be a grayscale. The dimension of the input array was returned using another function. If this was the condition, all input channels were initialized with the same values. But, in other conditions, the channel had to be defined. Then, the colors were applied to the complemented image using RGB 8-bits color channels. Finally, the red, green and blue channels were concatenated into 3-D arrays which produced the colored PN image. To get the desired output, the colorized function was used on the complemented image with the input resized image (512x512) in white [1 1 1] background. Fig. 13 shows the final outputs of colorization process.

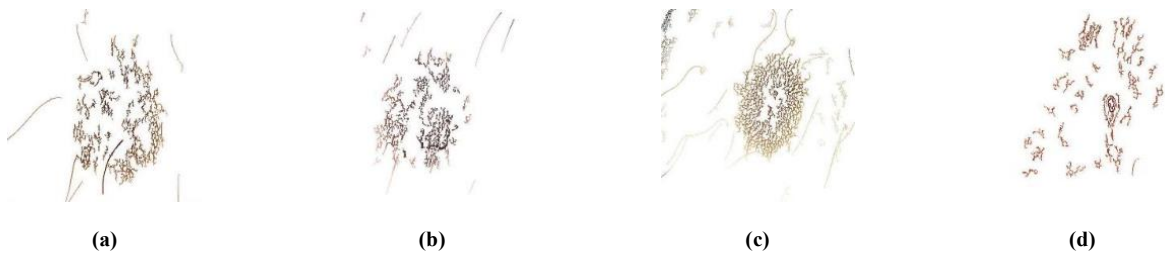

(a) (b) (c) (d)

**Fig. 13: (a), (b), (c) and (d) are the outputs of colorization process.**

### 3.4 PN classification

The pattern and color intensity are the primary differences between the APN and TPN. The line formations of the APN are the irregular dark-brown, gray, or black color [4, 34]. In contrast, the line of TPN forms with uniformly distributed light brown tint [5, 34]. Consequently, we implemented several machine learning-based classifiers to categorize them into two classes. We trained the SVM [35], ANN [36], CNN [37], and BoF [17] on the PN datasets. Among these four, we reported the best two, namely, the CNN and BoF. To demonstrate the effectiveness and preciseness of the imaging mask on identifying PN type, we used two datasets in the training process. First, we trained the CNN on the PH2 (lesion images containing PN) and on the new dataset (only PN images). Next, we trained the BoF on both datasets.

### 3.4.1 Classifier: CNN

Convolutional Neural Networks (CNNs) are essential tools for deep learning, particularly suited for medical image recognition. However, CNNs typically require a large amount of training data to perform effectively. In this study, we only had 200 images to classify pigment network (PN) into atypical and typical classes. To address this limitation, we designed a simple and effective CNN specifically developed for the PH2 dataset. Inspired by LeNet [38], CNN [39, 40], and DCNN [41], the proposed CNN architecture includes 2 convolutional layers and 2 batch normalization layers. We divided the dataset into training and validation sets, with the training set comprising 80% and the validation set 20% of the 200 images for each label. In the image input layer, we specified an image size of 280-by-280-by-3, corresponding to the height, width, and number of color channels (RGB images have 3 color channels). In the convolutional layers, we used filters with a kernel size of 5-by-5. The number of filters determined the number of feature maps, and we added padding to ensure the spatial output size matched the input size using the 'same' padding. Batch normalization layers were included between convolutional layers and activation functions to normalize the activations and gradients, making network training more efficient and reducing sensitivity to network initialization. Table 1 shows the architecture of the proposed CNN.

**Table 1**
The architecture of proposed CNN to classify PN on PH2 dataset.

| No. | Layer Type | Kernel Size* | No. of Filters/Neurons* | Dilation Factor* | Padding* | Stride* |
|---|---|---|---|---|---|---|
| 1 | Image/Input | - | - | - | - | - |
| 2 | Convolution | 5x5 | 8 | 3x3 | same | 1x1 |
| 3 | Normalization | - | - | - | - | - |
| 4 | Activation function* | - | 8 | - | - | - |
| 5 | Max Pooling | 5x5 | - | - | same | 2x2 |
| 6 | Convolution | 3x3 | 16 | 2x2 | same | 3x3 |
| 7 | Normalization | - | - | - | - | - |
| 8 | Activation function* | - | 16 | - | - | - |
| 9 | Fully connected | - | - | - | - | - |
| 10 | Softmax | - | - | - | - | - |
| 11 | Classification/Output | - | - | - | - | - |

*The parameters with (*) were selected by "Grid-Search" method to optimize the network's performance.*

We employed Clipped Rectified Linear Unit (Clipped-ReLU) as the activation function, as an ablation study showed it produced the most effective results for small datasets. The max-pooling layer, with a pool size of 2-by-2, was used to return the maximum values from rectangular regions of the input. The final fully connected layer combined the features to classify the images, with an output size of 2, corresponding to the two classes. The softmax layer normalized the outputs of the fully connected layer, producing positive numbers that sum to one, which could be interpreted as classification probabilities by the classification layer. The final classification layer used these probabilities to assign each input to one of the mutually exclusive classes and compute the loss.

After defining the CNN structure, we specified the training options. The CNN was trained using stochastic gradient descent with momentum with an initial learning rate of 0.01. We set the maximum number of epochs to 250. An epoch is a full training cycle on the entire training dataset. The network accuracy during training was monitored by specifying validation data and validation frequency. We shuffled the data in every epoch with 25 validation frequencies. The network was trained on the training data and calculated the accuracy of the validation data at regular intervals during training. The validation data was not used to update the network weights. Since the CNN was trained using the architecture defined by layers, each layer of the CNN produces a response, or activation to an input image. To visualize the learned-features of different layers, we applied "Deep Dream Image" [42] algorithm. In Fig. 14, important layers learned-features for both PH2 and new dataset are shown in (a) and (b).

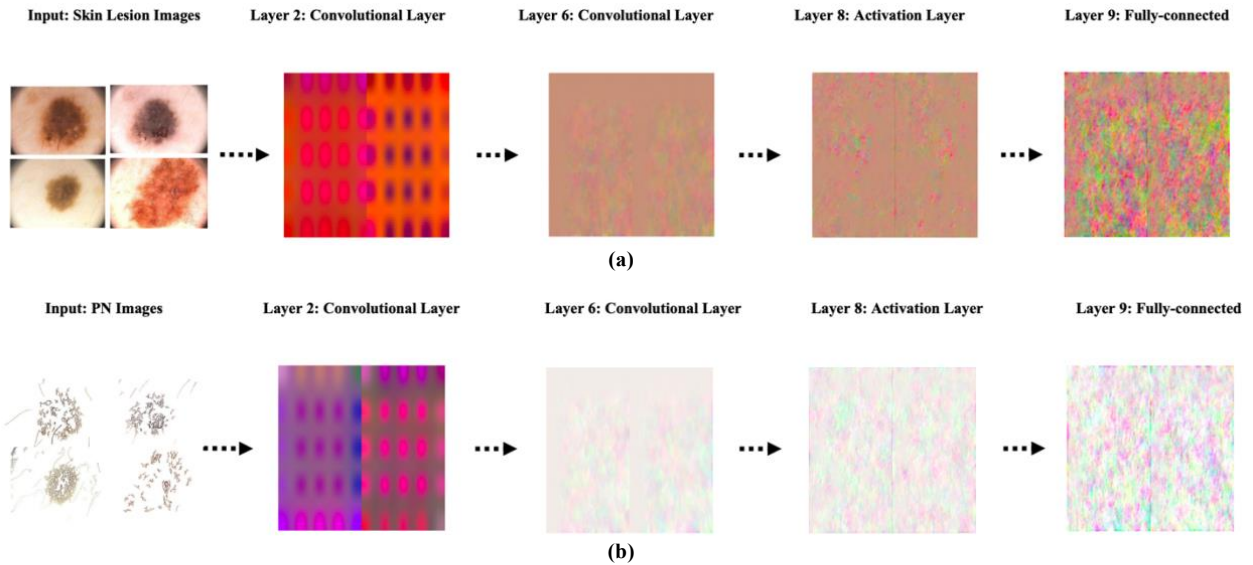

**Fig. 14: Important layers learned-features of CNN, (a) is for PH2 and (b) is for the new dataset.**

### 3.4.2 Classifier: BoF

We implemented the BoF approach to classify PN. This approach is also known as bag of words. The process of providing a category name to an image under test is known as visual image categorization. Images portraying just about any form of thing can be found in categories. We divided the data into two categories: training and validation. We chose 80% of the images from each batch for training and the remaining 20% for validation. To avoid bias in the results, we applied a split random. BoF is a natural language processing technique that has been extended to computer vision. We created a "vocabulary" of speeded up robust features (SURF) [43] representative of each image category because images do not contain discrete words. The SURF is a fast and robust algorithm for local, similarity invariant representation and comparison of images. This was done with a single call to the bagOfFeatures function, which extracts the SURF features from each image through all categories and creates the visual vocabulary by lowering the number of features using *K*-means clustering to quantize feature space.

The bagOfFeatures object also includes an encode function that counts the number of visual word occurrences in an image, resulting in a histogram—a new and simplified representation of the image. Fig. 15 shows examples of histogram images from the PH2 and new datasets.

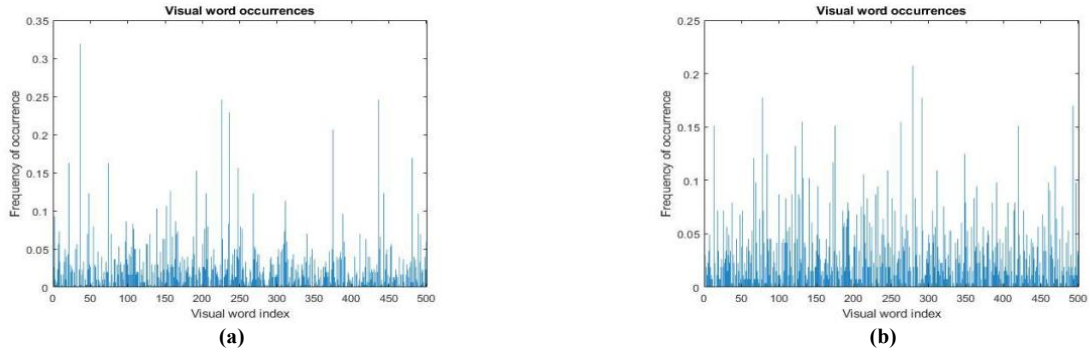

Fig. 15: One of the histogram images from PH2 is (a), and from the new dataset is (b).

This histogram was used to train a classifier as well as for the image categorization. It converted an image to a feature vector. The category classifier function called for the encoded training images from each category to be input for a classifier training procedure. The above function used the input bagOfFeatures objects encode method to generate feature vectors for each image category in the training set. An evaluation was performed upon completing the training process. This was followed by a sanity check with the training set, which yielded a near-perfect CM. The classifier was then evaluated on the validation set, which used datasets not utilized during the training process. The CM was returned by default by the evaluate function, which was a useful first indicator of how well the classifier was functioning.

## 4. RESULTS AND DISCUSSION
*4.1 Evaluation of proposed directional imaging algorithm on different datasets*
**PH2:** The success rate of directional imaging approaches was 96.00% for 200 images of PH2. Table 2 shows the success rate of the filter in two different PN classes. Our imaging filters could not detect 4 out of 83 images from the TPN and 4 out of 117 images from the APN because of the different pixel intensity of these images. Fig. 16 shows some of these 8 images. To detect PN using the same directional filter, a possible strategy was to reduce the threshold level of pixels (i.e., 0.001 to 0.011). Initially, by storing this 96% images from a total of 192 images out of 200 images, we created a new dataset with the only features of PN. Later, to extract PN, the pixel threshold level was changed individually for those 8 undetectable images (4 from APN and 4 from TPN). Finally, the PNs were isolated from them and stored in the new dataset with the rest of the 192 images. We used this new dataset (which contains a total of 200 TPN and APN images) for training different classifiers.

**Table 2**
The success rate of directional imaging filters in two different classes.

| Network type | Not detected | Detected |
|---|---|---|
| TPN | 4.82% | 95.18% |
| APN | 3.41% | 96.59% |

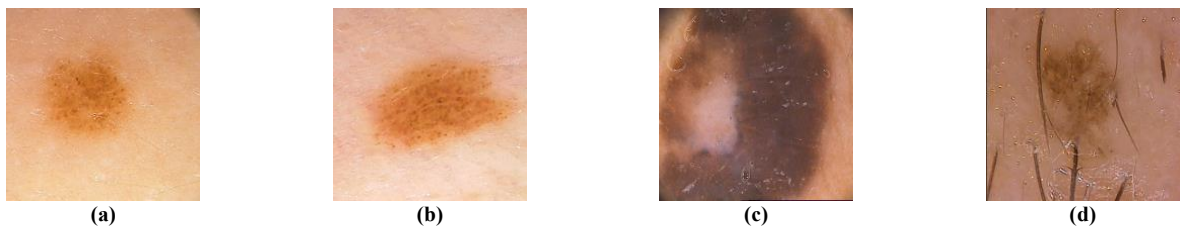

Fig. 16: Undetectable images (a) and (b) are from the TPN, and undetectable images (c) and (d) are from the APN.

**ISIC:** To show the flexibility of this imaging algorithm, we applied it on several dermoscopic images of different origins such as the ISIC 2019 and 2020. In Fig. 17, the input images of different sources and its corresponding output images (after applying the proposed imaging algorithm) are presented.

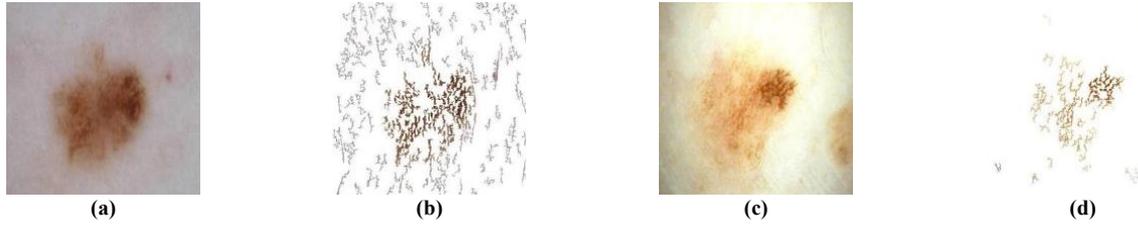

(a)　　　　　　　　(b)　　　　　　　　(c)　　　　　　　　(d)

**Fig. 17: (a) is input from ISIC 2019 and (b) is its output, and (c) is input from ISIC 2020 and (d) is its output.**

**Med-Node:** A non-dermoscopic dataset Med-Node was used to demonstrate the success of directional imaging algorithm for detecting and isolating PN. All the images of this dataset are macroscopic (a photograph or digital image taken through a microscope or similar device to show a magnified image of an object) images with Melanoma and Naevus lesion. In Fig. 18, the input images of non-dermoscopic source and its corresponding output images (after applying the proposed imaging algorithm) are presented.

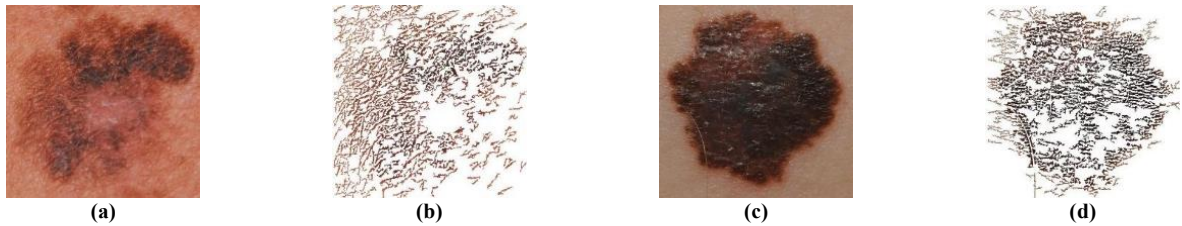

(a)　　　　　　　　(b)　　　　　　　　(c)　　　　　　　　(d)

**Fig. 18: (a) and (c) are inputs from MED-NODE where (b) and (d) are their outputs.**

The success rate of proposed filters on the ISIC and Med-Node datasets cannot be calculated similarly to the PH2 dataset because neither dataset contains annotated information on the PN.

*4.2 Evaluation of proposed classifiers: CNN and BoF*

In Fig. 19, the Confusion Matrices (CMs) for two different datasets are shown after predicting the labels of the validation data using the proposed CNN (trained) along with the final validation accuracy calculation. Accuracy is the fraction of labels that the network predicts correctly.

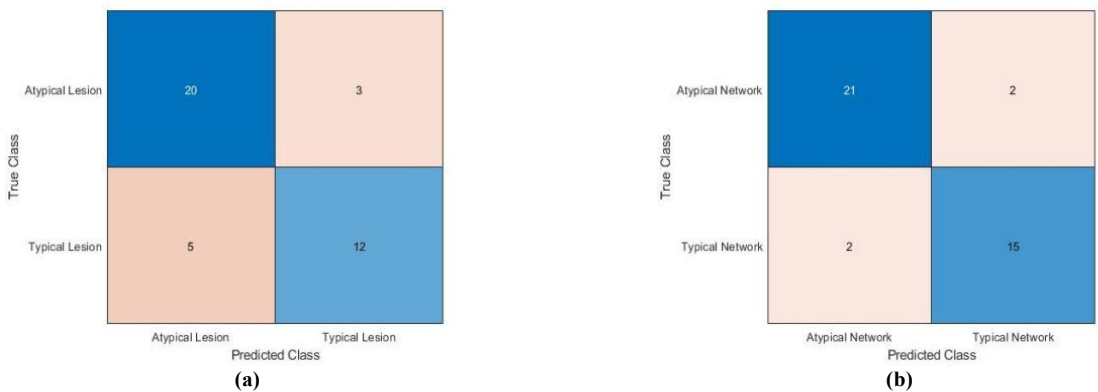

(a)　　　　　　　　　　　　　　　　　　(b)

**Fig. 19: The CM, (a) is on PH2 and (b) is on new dataset using CNN.**

We evaluated the CNN classifier's result using different evaluation metrics such as sensitivity, specificity, precision and accuracy for the CM (a) and (b). These evaluation metrics are computed as follows-

$$\text{Sensitivity, } SE = \frac{TP}{TP+FN} \quad (2)$$

$$\text{Specificity, } SP = \frac{TN}{FP+TN} \quad (3)$$

$$\text{Precision, } PR = \frac{TP}{TP+FP} \quad (4)$$

$$\text{Accuracy, } AC = \frac{TP+TN}{P+N} \quad (5)$$

Here, TP is true positive number, FN is false negative number, FP is false positive number, and TN is true negative number of the predicted labels of the validation data. The result of the evaluation is shown in Table 3 after applying Eq. (2) to Eq. (5) on Fig. 19.

**Table 3**
The result of evaluation metrics on two different datasets for CNN classifier.

| Datasets | SE | SP | PR | AC |
|---|---|---|---|---|
| PH2 | 0.80 | 0.78 | 0.80 | 0.80 |
| New | 0.90 | 0.89 | 0.90 | 0.90 |

After analyzing Table 3, it was found that the CNN classifier performs better on the PN-based (new) dataset than the lesion-based (PH2) dataset. In Fig. 20, the CMs for two different datasets are shown after predicting the labels of the validation set using the BoF trained classifier. The final evaluation result of this classifier is presented in Table 4 after applying Eq. (2) to Eq. (5) on Fig. 20.

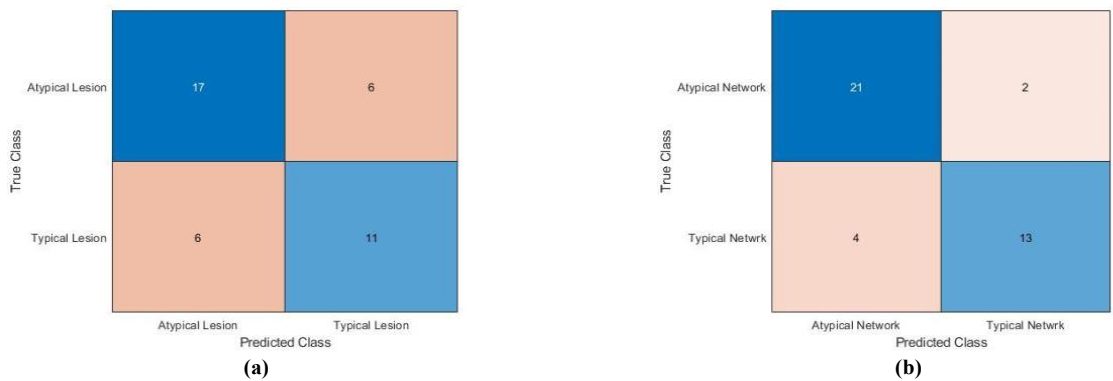

**Fig. 20: The CM, (a) is on PH2 and (b) is on the new dataset using BoF.**

**Table 4**
The result of evaluation metrics (%) on two different datasets for BoF classifier.

| Datasets | SE | SP | PR | AC |
|---|---|---|---|---|
| PH2 | 0.70 | 0.69 | 0.70 | 0.70 |
| New | 0.85 | 0.82 | 0.85 | 0.85 |

In Table 4, the BoF performs better same as the CNN on the new dataset than PH2. This indicates using new dataset helps to achieve better performance to classify PN. By comparing Table 3 and Table 4, we find that CNN is a better classifier than BoF for PN categorization. In Fig. 21, the ROC (receiver operating characteristic) curves of these two classifiers on two datasets are shown. The area under the ROC curve (AUC) for CNN and BoF are 0.76 and 0.69 (on PH2); 0.84 and 0.80 (on the new dataset).

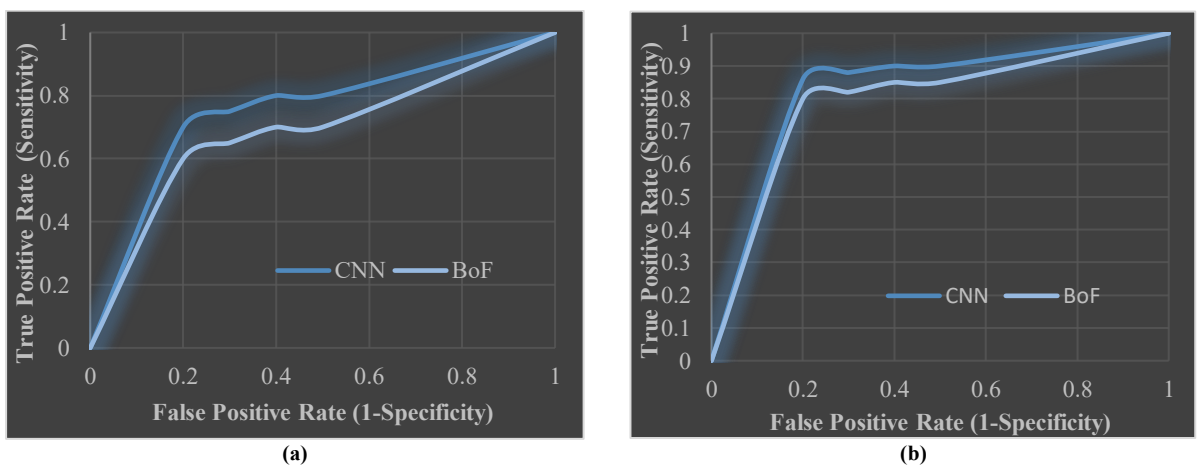

**Fig. 21. (a) ROC curves on PH2, and (b) ROC curves on new dataset for CNN and BoF respectively.**

*4.3 A comparative study*

To demonstrate the success of proposed PN identification approach over the state-of-art methods, the three existing methods were analyzed. For comparison process, we included the reported result of two layers feed-forward ANN proposed by Pathan et al. Then, we considered the outcome of the Adaboost classifier reported by Barata et al. In both cases, the experimental approaches were performed on the 200 dermoscopic images of PH2. Without implementing their approaches, we performed a direct comparison between their results and our proposed approach.

Finally, the proposed method was analyzed against the naïve Bayes classifier suggested by Kropidlowski et al. They implemented their own algorithm (based on region histogram, co-occurrence matrix, run-length matrix, gradient matrix, and autoregression model) on their dataset that contains only 44 PN images (19 APN and 25 TPN). For the sake of fair comparison, we implemented their algorithm on PH2 and evaluated the output. The results obtained indicate the superiority of the proposed approach over the other three approaches as shown in Table 5.

**Table 5**
Performance (%) comparison between state-of-the-art and the proposed methods for identifying PN on PH2.

| Approaches | SE | SP | PR | AC |
|---|---|---|---|---|
| ANN [19] | 84.6 | 88.7 | 86.7 | 86.7 |
| AdaBoost [21] | **91.1** | 82.1 | 86.2 | 86.2 |
| Naïve Bayes [23] | 85.0 | 88.5 | 85.0 | 87.0 |
| Proposed approach | 90.0 | **89.0** | 90.0 | 90.0 |

**5. CONCLUSION**

Despite significant achievements in this research, some limitations exist. The experiment was conducted using only the PH2 dataset, the sole publicly available dataset annotated with the ground truth of PN type. Future work could extend this research by acquiring another dataset like PH2 and applying the proposed method to it. Additionally, while the PN was categorized into typical and atypical classes, future work could refine the atypical category by splitting it into prominent APN and delicate APN based on color intensity and pattern. This research focused on APN, a marker of unhealthy skin lesions. Future research could also incorporate other dermatological features, such as asymmetrical shape, blue-white veils, and colors of skin lesions, to enhance melanoma diagnosis.

To identify PN from dermoscopic images, a directed imaging algorithm incorporating PCA, and contrast-enhanced techniques was used. The initial success rate of this method was 96%, which increased to 100% with pixel intensity modification. Utilizing this algorithm on PH2, a new PN-based dataset was developed to train two classifiers, CNN and BoF, with CNN demonstrating superior performance in categorizing the networks. The findings from this study, including the detection and classification rate of PN, outperform existing approaches in the literature. The proposed technique's flexibility allows for slight modifications to the directional filters, enabling the detection of other dermoscopic structures such as dots-globules, streaks, and lesion colors. Future research could explore integrating additional dermatological features into the proposed CNN to enhance diagnostic accuracy for conditions like melanoma. Incorporating larger and more diverse datasets could improve the proposed CNN's robustness and generalizability. The continuous development of advanced machine learning techniques and their application to dermoscopic imaging holds great potential for improving early detection and diagnosis of skin lesions.

**CRediT authorship contribution statement**
**M. A. Rasel:** Conceptualization, Methodology, Software, Validation, Formal analysis, Data Curation, Writing – Original Draft, Visualization. **Sameem Abdul Kareem**: Conceptualization, Formal analysis, Investigation, Writing - Review & Editing, Supervision. ***Unaizah Obaidellah:*** Conceptualization, Writing- Reviewing and Editing, Supervision, Project administration.

**Declaration of competing interest**
The authors declare that they have no known competing financial interests or personal relationships that could have appeared to influence the work reported in this paper.

**Data availability**
Data will be made available on request.